\documentclass[reprint,twocolumn,showpacs,amssymb,amsmath,aps,pra]{revtex4}
\newcommand{\half}{\mbox{$\frac{1}{2}$}}
\usepackage{graphics}
\usepackage{bm}
\begin{document}
\title{Separability conditions and limit temperatures for entanglement
detection in two qubit Heisenberg $XYZ$ models}
\author{N. Canosa, R. Rossignoli}
\affiliation{Departamento de F\'{\i}sica, Universidad Nacional de La Plata,
C.C.67, La Plata (1900), Argentina}
\begin{abstract}

We examine the entanglement of general mixed states of a two qubit Heisenberg
$XYZ$ chain in the presence of a magnetic field, and its detection by means of
different criteria. Both the exact separability conditions and the weaker
conditions implied by the disorder and the von Neumann entropic criteria are
analyzed. The ensuing limit temperatures for entanglement in thermal states of
different $XYZ$ models are then examined and compared with the limit
temperature of the symmetry-breaking solution in a mean field type
approximation. The latter, though generally lower, can also be higher than the
exact limit temperature for entanglement in certain cases, indicating that
symmetry-breaking does not necessarily entail entanglement. The reentry of
entanglement for increasing temperatures is also discussed.
\end{abstract}
 \maketitle
\section{Introduction}
Entanglement is one of the most distinctive features of quantum mechanics,
representing the ability of composite quantum systems to exhibit correlations
which have no classical analogue. Recognized already by Schr\"odinger
\cite{S.35}, it has recently become the object of intensive research due to the
key role it plays in the field of quantum information
\cite{Ek.91,Be.93,Di.95,Be.00,NC.00}. Rigorously, a mixed state $\rho$ of a
bipartite system is said to be {\it separable} or {\it classically correlated}
\cite{W.89} if it can be expressed as a convex combination of uncorrelated
densities, i.e., $\rho=\sum_\nu q_\nu\rho^\nu_A\otimes\rho^\nu_B$, where
$\rho^\nu_A$, $\rho^\nu_B$ are mixed states of each subsystem and $q_\nu$ are
{\it non-negative} numbers. Otherwise, $\rho$ is {\it entangled} or
inseparable. When separable, $\rho$ satisfies all Bell inequalities as well as
other properties characteristic of classical systems.

A pure state $\rho=|\Phi\rangle\langle\Phi|$ is separable just for tensor
product states $|\Phi\rangle=|\phi_A\rangle|\phi_B\rangle$, but in the case of
mixed states, like thermal states $\rho\propto\exp[-H/T]$, with $H$ the system
Hamiltonian, it is in general much more difficult to determine whether $\rho$
is separable or not. Only in special cases, like a two-qubit or qubit+qutrit
system, simple necessary and sufficient conditions for separability are known
\cite{P.96,HHH.96}. Moreover, the entanglement of formation of a mixed state
\cite{Be.96} has been explicitly quantified only for a two-qubit system
\cite{W.98}. Nonetheless, it is known that any mixed state becomes separable if
it is sufficiently close to the fully mixed state \cite{ZHSL.98,B.99}. For
thermal states of finite systems, this implies that a {\it finite} limit
temperature for entanglement \cite{N.98}, $T_e$, will always exist such that
$\rho$ becomes separable $\forall$ $T\geq T_e$. It is then interesting to
analyze if it is possible to estimate this temperature with simple separability
criteria, and how it is related to the critical temperature $T_c$ of the {\it
symmetry-breaking} solution in a mean field type approximation, which is the
conventional starting point for describing interacting many-body systems. Such
solutions (i.e., like deformed or superconducting) normally reflect the
presence of strong correlations and collective behavior.

The aim of this work is to examine these issues in a simple yet non-trivial
model where the exact entanglement conditions and quantification can be easily
obtained. For this purpose, we will consider a system of two qubits interacting
through a Heisenberg $XYZ$ Hamiltonian \cite{K.93} in the presence of an
external magnetic field. Interest in this model stems from the potential use of
Heisenberg spin chains for gate operations in solid state quantum computers
\cite{I.99,RB.01}. The pairwise entanglement of thermal states of isotropic
\cite{Ar.01,W.02} and anisotropic XY \cite{W.01,KS.02,ON.02} Heisenberg models
have accordingly been recently studied, and several interesting features have
appeared already in the two-qubit case \cite{KS.02}, like the possibility of
entanglement reentry for increasing temperatures or magnetic fields.

We will first review the exact separability conditions for {\it general}
mixtures of the eigenstates of arbitrary $XYZ$ Hamiltonians, examining in
particular thermal states and the possibility of entanglement reentry. We will
also analyze the weaker conditions provided by the {\it disorder} criterion
\cite{NK.01}, which is the strongest one based just on the spectrum of $\rho$
and one of its reductions, and is hence more easy to implement in general than
other criteria. Violation of the disorder conditions also ensures
distillability \cite{H.03}. These conditions are here {\it exact} in the
absence of a magnetic field. Although the disorder criterion admits a
generalized entropic formulation \cite{RC.02}, it is stronger than the von
Neumann entropic criterion \cite{HH.96}, based on the same information, whose
predictions will also be analyzed. The ensuing exact and approximate limit
temperatures for entanglement in thermal states of different XYZ models will
then be examined.

Finally, we will discuss the mean field (i.e., independent qubit) approximation
for thermal states, with the aim of comparing the previous limit temperatures
with the corresponding mean field critical temperature $T_c$. It will be shown,
remarkably, that for $T>0$, {\it symmetry-breaking  is not necessarily a
signature of entanglement}, so that $T_c$ may be {\it higher} than $T_e$,
although it is usually lower. The model and methods are described in sec.\ II,
while three different examples are analyzed in detail in sec.\ III. Conclusions
are finally drawn in IV.

\section{Formalism}
\subsection{Model and separability conditions}
We will consider a Heisenberg $XYZ$ chain \cite{K.93} for two qubits in an
external magnetic field $b$ along the $z$ axis. Denoting with
$\bm{S}=\bm{s}^A+\bm{s}^B$ the total spin of the system, the corresponding
Hamiltonian can be written as
 \begin{subequations}\label{H}\begin{eqnarray}
H&=&bS_z-2\sum_{i=x,y,z}v_is_i^As_i^B \label{H1}\\
 &=&H_z-v_+(S_x^2+S_y^2-1)-v_-(S_x^2-S_y^2)\,,\label{H2}
 \end{eqnarray}\end{subequations}
where $H_z=bS_z-v_z(S_z^2-1/2)$ and $v_{\pm}=(v_x\pm v_y)/2$. The ferromagnetic
(antiferromagnetic) case corresponds to $v_i\geq 0$ ($\leq 0$), and the
standard $XY$ model to $v_z=0$. Its normalized eigenstates
$H|\Phi_j\rangle=E_j|\Phi_j\rangle$ are given by
 \begin{eqnarray}   &&\begin{array}{lcll}
 |\Phi_{0,3}\rangle&=&\frac{|+-\rangle\mp|-+\rangle}{\sqrt{2}}\,,
 &\;E_{0,3}=\half v_z\pm v_+\,,\\
 |\Phi_{1,2}\rangle&=&\frac{u_{\pm}|\!++\rangle\mp
 u_{\mp}|\!--\rangle}{\sqrt{2}}\,,&\;E_{1,2}= -\half v_z\pm \Delta\,,
 \end{array}\label{ba}
 \end{eqnarray}
with $\Delta=v_-\sqrt{1+b^2/v_-^2}$, $u_{\pm}=\sqrt{1\pm b/\Delta}$ and
$|\!\pm\pm\rangle\equiv|\pm\rangle|\pm\rangle$ the separable eigenstates of
$S_z$ (standard basis). The states $|\Phi_{0,3}\rangle$ are maximally
entangled, whereas $|\Phi_{1,2}\rangle$ are entangled for $v_{-}\neq 0$, with
concurrence $v_-/\Delta$ (see Appendix). They become maximally entangled for
$b=0$, in which case the set of states (\ref{ba}) is just the Bell basis.

We will first consider general statistical mixtures of the previous
eigenstates, which can be written as
 \begin{subequations}\label{4}\begin{eqnarray}
\rho&=&\sum_{j=0}^3p_j|\Phi_j\rangle\langle\Phi_j| \label{00}\\
&=&{\textstyle\frac{1}{4}}+\half\langle S_z\rangle S_z+ 4\sum_{i=x,y,z}\langle
s_i^As_i^B\rangle s_i^As_i^B,
 \label{gral}\end{eqnarray}\end{subequations}
where $p_j\geq 0$, $\sum_{j=0}^3p_j=1$ and
 \begin{equation}\begin{array}{lcl}
\langle S_z\rangle&=&{\frac{b}{\Delta}}(p_1-p_2)\,,\;\;
\langle s_z^As_z^B\rangle=\half(p_1+p_2-\half)\,,\\
 \langle s_{i}^As_{i}^B\rangle&=&{\textstyle\frac{1}{4}}
 [p_3-p_0\pm{\frac{v_-}{\Delta}}(p_2-p_1)]\,,\;i=x,y\,,
 \end{array}\label{S}\end{equation}
with $\langle O\rangle\equiv {\rm Tr}\rho O$. Eqs.\ (\ref{4}) comprise standard
thermal states as well as those arising in more general statistical
descriptions \cite{T.88,CR.02}, and represent the most general two-qubit state
with good permutational and phase flip symmetry $U=-e^{i\pi S_z}$ real in the
standard basis. The two-site density matrix of an $N$ qubit $XYZ$ chain with
cyclic boundary conditions is in fact also of this form \cite{ON.02}.

{\it Exact separability conditions.} For the state (\ref{4}), they can be most
easily determined with the Peres criterion \cite{P.96}, sufficient for two
qubits \cite{HHH.96}, and can be cast  as
\begin{subequations}\label{i}\begin{eqnarray}
 \!\!\!\!{\textstyle\frac{v_-}{\Delta}}|p_2-p_1|&\leq& p_0+p_3\,,\label{i12}\\
 \!\!\!\!|p_3-p_0|&\leq&[(p_1+p_2)^2
 -{\textstyle\frac{b^2}{\Delta^2}}(p_2-p_1)^2]^{\frac{1}{2}}\,,\label{i03}
 \end{eqnarray}\end{subequations}
or, in terms of the averages
 $\langle S_i^2\rangle=2\langle s_i^As_i^B\rangle+1/2$, as
 \begin{subequations}\label{e}\begin{eqnarray}
 |\langle S_x^2-S_y^2\rangle|&\leq &\langle 1-S_z^2\rangle\,,\label{e1}\\
 |\langle S_x^2+S_y^2-1\rangle|&\leq&[\langle S_z^2\rangle^2-
 \langle S_z\rangle^2]^{\frac{1}{2}} \,,\label{e2}
 \end{eqnarray}\end{subequations}
imposing bounds on the averages of the last two terms in (\ref{H2}). If $\rho$
is entangled, only one of Eqs.\ (\ref{i}) is violated, and its concurrence is
given precisely by the difference between the left and right hand sides of the
broken inequality (see Appendix). The entanglement arises essentially from one
of the states $|\Phi_{1,2}\rangle$ ($|\Phi_{0,3}\rangle$) if (\ref{i12})
[(\ref{i03})] is broken. Eqs.\ (\ref{i}) are always satisfied if $|p_j-1/4|\leq
(4\sqrt{2})^{-1}$ $\forall$ $j$, i.e., if $\rho$ is sufficiently close to the
fully mixed state. If $b=0$, $\rho$ is diagonal in the Bell basis and Eqs.\
(\ref{i}) reduce accordingly to $p_j\leq 1/2$ $\forall$ $j$ \cite{HH.96}, while
Eqs.\ (\ref{e}) to $1\leq \langle S^2\rangle\leq 1+2\langle S_i^2\rangle$ for
$i=x,y,z$, as $\langle S_z\rangle=0$.

{\it Disorder and entropic separability conditions}. The disorder criterion
\cite{NK.01} states that if $\rho$ is separable, $\rho$ is {\it majorized} by
the reduced densities $\rho_{A,B}\equiv {\rm Tr}_{B,A}\,\rho$, which means that
$\rho$ is  {{\it more mixed} (i.e., {\it disordered}) than  $\rho_A$, $\rho_B$.
In a two qubit system, this implies that the largest eigenvalue of $\rho$
should not exceed that of $\rho_A$ and $\rho_B$, which is in general a
necessary condition that becomes {\it sufficient} when $\rho$ is pure or {\it
 diagonal} in the Bell basis \cite{NK.01,RC.02}.

For the state (\ref{4}), $\rho_{\alpha}=1/2+\langle S_z\rangle s_z^{\alpha}$
for $\alpha=A,B$, and the disorder criterion leads to the inequalities
 \begin{equation}
 p_j\leq \half[1+|{\textstyle\frac{b}{\Delta}}(p_2-p_1)|]\,,
 \;j=0.\ldots,3\,,\label{dis}
 \end{equation}
which in terms of total spin averages can be recast as
 \begin{subequations}\label{d}\begin{eqnarray}
&& \!\!\!\!\!\!\!\!\!\!\!\!\!\!\!\!\!\!\!
|\langle S_x^2-S_y^2\rangle|\leq
\langle 1-S_z^2\rangle[1+2|\langle S_z\rangle|/\langle 1-
S_z^2\rangle]^{\frac{1}{2}}\,,\label{d1}\\
 &&\!\!\!\!\!\!\!\!\!\!\!\!\!\!\!\!\!\!\!
 |\langle S_x^2+S_y^2-1\rangle|\leq\langle S_z^2\rangle
 +|\langle S_z\rangle|\label{d2}\,.
 \end{eqnarray}\end{subequations}
Eqs.\ (\ref{dis}) or (\ref{d}) are clearly less stringent in general than Eqs.\
(\ref{i}) or (\ref{e}), but become {\it exact} for $b=0$ ($\langle
S_z\rangle=0$), i.e., when $\rho$ is diagonal in the Bell basis.

The standard entropic criterion \cite{HH.96}, based on the von Neumann entropy
$S_2(\rho)=-{\rm Tr}\,\rho\log_2\rho$, states that if $\rho$ is separable,
$S_2(\rho)\geq S_2(\rho_\alpha)$ for $\alpha=A,B$. Although exact for pure
states (in which case $S_2(\rho)=0$ and $S_2(\rho_A)=S_2(\rho_B)$ is just the
entanglement of $\rho$ \cite{W.98}), for mixed states it is in general weaker
than the disorder criterion \cite{RC.02}, except when both $\rho$ and
$\rho_{\alpha}$ have rank two. Fig.\ \ref{f1} depicts, for $p_1=0$ and
$b/v_-=1$, the regions where the state (\ref{4}) is entangled and where
entanglement is detected by the disorder and the standard entropic criteria.

\begin{figure}[t]
 \centerline{\scalebox{0.38}{\includegraphics{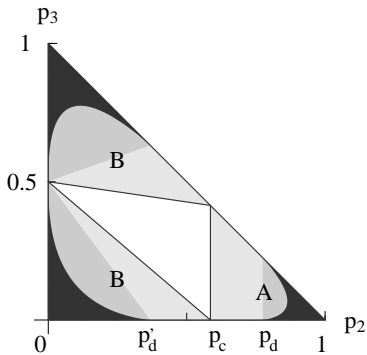}}}
\vspace*{-.3cm}

\caption{Range of values of $p_2$ and $p_3$ where the state (\ref{4}) is
entangled (shaded sectors), for $p_1=0$, $b/v_-=1$. Eq.\ (\ref{i12}) is broken
in sector $A$ while Eq.\ (\ref{i03}) in sectors $B$. Entanglement is detected
by the disorder criterion in the black and dark gray sectors, and by the von
Neumann entropic criterion just in the black sectors. They both coincide at the
border of the triangle, where $\rho$ has rank 2, and are exact for $p_2=0$,
where $\rho$ is in addition diagonal in the Bell basis.} \label{f1}
\vspace*{-.45cm}
\end{figure}

{\it Standard thermal state and entanglement reentry}. For
 \begin{equation}
\rho=\exp[-\beta H]/{\rm Tr}\exp[-\beta H]\,,\;\;\;\;\beta\equiv 1/T>0\,,
 \label{1}
 \end{equation}
i.e., $p_j\propto e^{-\beta E_j}$ in (\ref{00}),  Eqs.\ (\ref{i}) become
 \begin{subequations}\label{t}\begin{eqnarray}
{\textstyle\frac{v_-}{\Delta}e^{\beta v_z}\sinh|\beta\Delta|}
 &\leq&\cosh(\beta v_+)\,,\label{t1}\\
 e^{-\beta v_z}\sinh|\beta v_+|&\leq&{\textstyle
 [1+\frac{v_-^2}{\Delta^2}\sinh^2(\beta\Delta)]^{\frac{1}{2}}}\,,
 \label{t2} \end{eqnarray}\end{subequations}
and determine a finite limit temperature for entanglement $T_e$, such that they
are satisfied $\forall$ $T\geq T_e$. Nonetheless, entanglement, as measured by
the entanglement of formation or concurrence, may not be a decreasing function
of $T$ for $T<T_e$ when the ground state is less entangled than the first
excited state \cite{Ar.01,W.02,W.01,KS.02}, and even entanglement vanishing
plus reentry may occur \cite{KS.02}, as discussed below. Note that the spectrum
of $H$, and hence Eqs.\ (\ref{t}), do not depend on the signs of $b$ and
$v_{\pm}$.

Let us consider for instance a mixture of $|\Phi_2\rangle$ and $|\Phi_3\rangle$
($p_0=p_1=0$ in (\ref{00})), which corresponds to the outer border in fig.\
\ref{f1}. This state is separable {\it just for} $p_2=p_c\equiv
(1+v_-/\Delta)^{-1}\geq 1/2$, with Eq.\ (\ref{i12}) [(\ref{i03})] broken for
$p_2>p_c$ ($p_2<p_c$). Its concurrence is $C(\rho)=|p_2/p_c-1|$. The state
(\ref{1}) will approximately be of this form for low $T$ if $E_2$ and $E_3$ are
sufficiently close and well below the remaining levels. Hence, if $E_2<E_3$,
$C(\rho)$ will initially decrease as $T$ increases from zero, {\it vanishing}
at the temperature
\begin{equation}
 T_r=(E_3-E_2)/\ln[\Delta/v_-]\,,\label{Tr}
 \end{equation}
where $p_2=p_c$, but will exhibit a {\it reentry} for $T>T_r$, with Eq.\
(\ref{t1}) [(\ref{t2})] broken for $T<T_r$ ($T_r<T<T_e$). Due to the remaining
levels, $C(\rho)$ will actually vanish in a small but {\it finite} temperature
interval around $T_r$ (see case 3 in sec.\ III). When $|\Phi_2\rangle$ becomes
separable ($v_-/b\rightarrow 0$), $p_c\rightarrow 1$ and $T_r\rightarrow 0$,
whereas when it becomes maximally entangled ($b/v_-\rightarrow 0$),
$p_c\rightarrow 1/2$ and $T_r\rightarrow\infty$, so that no reentry takes place
in this limit. In constrast, if $E_3<E_2$, $p_2<1/2\leq p_c$ $\forall$ $T\geq
0$ and no reentry or enhancement of $C(\rho)$ can take place. Nor can it occur
for a mixture of $|\Phi_0\rangle$ and $|\Phi_3\rangle$ or $|\Phi_1\rangle$ and
$|\Phi_2\rangle$, since they are separable just for equal weights, as seen from
Eqs.\ (\ref{i}).

For the state (\ref{1}), the disorder conditions (\ref{dis}) become
 \begin{subequations}\label{dt}\begin{eqnarray}
&& \!\!\!\!\!\!\!\!\!\!\!\!\!\!\!\!\!\!\!
({\textstyle 1-|\frac{b}{\Delta}|})e^{\beta v_z} \sinh|\beta \Delta|\leq
       \cosh(\beta v_+)\label{dis1}\,,\\
&& \!\!\!\!\!\!\!\!\!\!\!\!\!\!\!\!\!\!\!
e^{-\beta v_z}\sinh|\beta v_+|\leq
 \cosh(\beta\Delta)+{\textstyle|\frac{b}{\Delta}}
 |\sinh|\beta\Delta|\label{dis2}\,,
 \end{eqnarray}\end{subequations}
and lead to a lower limit temperature for entanglement detection, $T^d_e\leq
T_e$, with $T_e^d=T_e$ just for $b=0$.  The entropic criterion leads to an even
lower limit temperature $T_e^s\leq T_e^d$. The reentry effect {\it cannot be
detected} by the disorder (and hence by the entropic) criterion. Violation of
Eqs.\ (\ref{dis}) requires $p_j>1/2$ for some $j$, so that in the thermal case
just the entanglement arising from the {\it ground} state can be detected. For
a mixture of $|\Phi_2\rangle$ and $|\Phi_3\rangle$, Eqs.\ (\ref{dis}) are
broken just for $p_2>p_d=(2-|b/\Delta|)^{-1}$ or $p_2<p'_d=(2+|b/\Delta|)^{-1}$
(see Fig.\ \ref{f1}), which does not allow to detect the reentry when $E_2<E_3$
since $p'_d\leq 1/2$.

\subsection{Symmetry-breaking mean field approximation}
The thermal state (\ref{1}) represents the density operator that minimizes the
free energy
 \begin{equation}
 F(\rho)\equiv\langle H\rangle-TS(\rho)=
 {\rm Tr}\rho[H+T\ln \rho]\,.\label{F}
 \end{equation}
In a finite temperature mean field or independent qubit approximation, Eq.\
(\ref{F}) is minimized among the subset of {\it uncorrelated} trial densities,
given in this case by
\begin{equation}
\rho_{\rm mf}=\rho_A\otimes\rho_B\,,\label{rmf}
\end{equation}
with arbitrary $\rho_A$, $\rho_B$, obtaining thus an upper bound to the minimum
free energy. The only way such an approximation can reflect entanglement is
through {\it symmetry breaking}: the optimum density that minimizes
$F(\rho_{\rm mf})$ may break some of the symmetries present in the Hamiltonian
$H$, and become degenerate. In these cases a {\it critical temperature} $T_c$
will exist such that the optimum density becomes symmetry conserving for $T\geq
T_c$. At $T=0$, {\it symmetry breaking implies entanglement} if the ground
state of $H$ is non-degenerate, since for pure states separability corresponds
to an uncorrelated density. However, this is not necessarily the case for
$T>0$, where symmetry-breaking just indicates, in principle, that the true
thermal state is not uncorrelated. On the other hand, entanglement does not
necessarily imply symmetry-breaking either, both at $T=0$ or $T>0$, as
correlations need to be in general sufficiently strong to induce a
symmetry-breaking mean field \cite{RS.80}.

The densities $\rho_\alpha$, $\alpha=A,B$, can be parameterized as
\begin{eqnarray}
\rho_{\alpha}&=&\frac{\exp[-\beta\bm{\lambda}^\alpha\cdot \bm{s}^\alpha]} {{\rm
Tr}\exp[-\beta\bm{\lambda}^\alpha\cdot \bm{s}^\alpha]} =\half+2\langle
\bm{s}^\alpha\rangle\cdot \bm{s}^\alpha\,,\\ \langle
\bm{s}^\alpha\rangle&=&{\rm Tr}\rho_{\alpha}\, \bm{s}^\alpha=
-\half\bm{\lambda}^\alpha\tanh[\half\beta|\bm{\lambda}^\alpha|]/
|\bm{\lambda}^\alpha|\,,\nonumber
\end{eqnarray}
so that Eq.\ (\ref{rmf}) corresponds to an approximate independent qubit
Hamiltonian $h=\sum_\alpha\bm{\lambda}^\alpha\cdot\bm{s}^\alpha$. Minimization
of $F(\rho_{\rm mf})$ with respect to $\bm{\lambda}^{\alpha}$ leads then to the
self-consistent equations (see for instance \cite{R.92})
 \begin{eqnarray}
 \lambda_i^\alpha&=&\frac{\partial\langle H\rangle_{\rm mf}}{\partial\langle
 s_i^\alpha\rangle}\,,\;\;i=x,y,z\,, \label{sc}
 \end{eqnarray}
where $\langle H\rangle_{\rm mf}={\rm Tr}\,\rho_{\rm mf}\,H$. A similar
equation obviously holds for the $n$ qubit case. In the case (\ref{H}),
$\langle H\rangle_{\rm mf}= b\langle S_z\rangle-2\sum_{i}v_i\langle
s_i^A\rangle\langle s_i^B\rangle$ and Eqs.\ (\ref{sc}) become
 \begin{equation}
 \lambda_i^{A,B}=b\delta_{iz}-2v_i\langle s_i^{B,A}\rangle\,.
 \end{equation}
Permutational symmetry will be broken if $\bm{\lambda}^A\neq \bm{\lambda}^B$,
and phase flip symmetry if $\lambda_x^\alpha\neq 0$ or $\lambda ^\alpha_y\neq
0$. The latter has to be broken in order to see any effect from the last two
interaction terms in (\ref{H2}) at the mean field level, since otherwise their
mean field averages vanish. In such a case the sign of one of the
$\lambda_x^\alpha$ (or $\lambda_y^\alpha$) remains undetermined, giving rise at
least to a two-fold degeneracy.

For instance, in the ferromagnetic case $v_i\geq 0$, $\langle H\rangle_{\rm
mf}$ is minimum for $\langle\bm{s}^A\rangle=\langle \bm{s}^B\rangle$ and
permutational symmetry needs not be broken. Hence,
$\bm{\lambda}^{A,B}=\bm{\lambda}$. Defining $v_M={\rm Max}[v_x,v_y]$,
$v_{m}={\rm Min}[v_x,v_y]$, a phase-flip symmetry breaking solution with
$|\lambda_M|\neq 0$ and $\lambda_m=0$ becomes feasible and provides the lowest
free energy if $v_M>v_z$ and $|b|<b_c\equiv v_M-v_z$, provided  $0\leq T<T_c$,
with
 \begin{equation}
 T_{c}=v_M\chi/\ln[\frac{1+\chi}{1-\chi}]\,,\;\;\;\chi\equiv |b|/b_c<1\,.
 \label{Temf}
 \end{equation}
$T_c$ decreases as $\chi$ increases, with $T_c\rightarrow 0$ for
$\chi\rightarrow 1$ and $T_c\approx\half v_M(1-\chi^ 2/3)$ for $\chi\ll 1$.
This solution is insensitive to $v_m$. As discussed in sec.\ III, $T_c$ is
usually lower than $T_e$, but can also be {\it higher}. For example, if $b=0$
and $v_x>v_y=v_z>0$, $T_c=v_x/2$, but the ensuing exact thermal state, diagonal
in the Bell basis, is {\it separable} $\forall\,T>0$ ($T_e=0$), as the ground
state is degenerate ($E_2=E_3-v_x/2$) and hence $p_j\leq 1/2$ $\forall$ $j,T$.

\section{Examples}
We now examine in detail the previous limit temperatures in three different
cases. We set in what follows $b\geq 0$, $v_{\pm}\geq 0$, since the concurrence
and limit temperatures are independent of their signs.

{\it 1)} $v_-=0$, $v_+>0$ ($XXZ$ model). The states $|\Phi_{1,2}\rangle$ are in
this case {\it separable}, with $\Delta=b$ in Eq.\ (\ref{ba}). Entanglement can
then only arise through the violation of Eq.\ (\ref{i03}),  i.e., Eq.\
(\ref{t2}) in the thermal case, which is now {\it independent} of the magnetic
field $b$. If $v_+>v_z$, the thermal state (\ref{1}) will then be entangled for
{\it any} $b$ if $T>0$, up to a limit temperature $T_e$ that is {\it
independent} of $b$. However, the ground state is $|\Phi_3\rangle$ if
$b<b_0\equiv v_+-v_z$ and $|\Phi_2\rangle$ if $b>b_0$, so that for $b>b_0$,
$\rho$ becomes entangled only at {\it finite temperature} $T>0$, in agreement
with Eq.\ (\ref{Tr}) ($T_r\rightarrow 0$ for $v_-/b\rightarrow 0$). On the
other hand, if $v_+<v_z$, no entanglement occurs at any temperature. These
features can be appreciated in Fig.\ \ref{f2} for $v_z=0$ ($XX$ model), where
$b_0=v_+$ and  \cite{W.01}
 \begin{equation}
 T_e=\alpha v_+\,,\;\; \alpha=1/\ln[1+\sqrt{2}]\approx 1.134\,. \label{tc1}
 \end{equation}
 \begin{figure}[t]
\vspace*{-2.6cm}

 \centerline{\scalebox{0.5}{\includegraphics{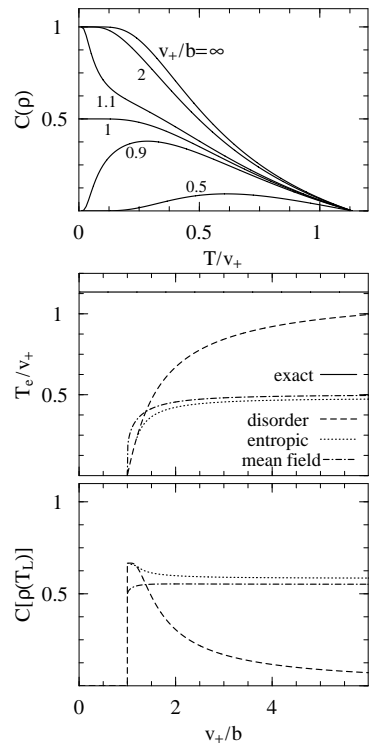}}}
 \vspace*{-2.8cm}

\caption{Top: The concurrence as a function of temperature for $v_-=v_z=0$ and
indicated values of $v_+/b$. Center: The corresponding exact limit temperature
for entanglement (constant) and the limit temperatures below which entanglement
is detected by the disorder and by the von Neumann entropic criterion. The
critical temperature for the symmetry-breaking mean field approximation is also
shown. Bottom: The concurrence at the previous limit temperatures.}
 \label{f2}
\end{figure}
The disorder criterion can now detect entanglement just through the violation
of Eq.\ (\ref{dis}) for $j=0,3$, i.e., Eq.\ (\ref{dis2}) in the thermal case,
which can occur only for $b<b_0$, i.e., when $|\Phi_3\rangle$ is the ground
state. The entanglement arising for $T>0$ when $b>b_0$ cannot be detected. In
addition, the limit temperature $T^d_e$ determined by (\ref{dis2}) will depend
on $b$, decreasing as $b$ increases and vanishing for $b\rightarrow b_0$. Its
behavior for $v_z=0$ is shown in the central panel of fig.\ \ref{f2}, where
$T^d_e\approx T_e[1-b/(\sqrt{2}v_+)]$ for $b\rightarrow 0$ while $T^d_e\approx
(v_+-b)/\ln 2$ for $b\rightarrow b_0=v_+$. Also shown is the concurrence at
$T=T^d_e$ (bottom panel), which is maximum at $b=b_c$ (where
$C[\rho(T^d_e)]\rightarrow 2/3$) and decreases as
$\approx\alpha^{-1}(\sqrt{2}-1)b/v_+$ for $b\rightarrow 0$. The limit
temperature $T_e^s$ of the entropic criterion is still lower. For $v_z=0$ and
$b\rightarrow 0$, $T^s_e\rightarrow 0.478v_+$, with $C[\rho(T_e^s)]\rightarrow
0.584$.

In this case the ground state critical field $b_0$ {\it coincides} with the
mean field critical field $b_c$. Hence, a stable symmetry breaking mean field
solution is here feasible just for $b<b_0$, with $T_c$ given by Eq.\
(\ref{Temf}) with $v_M=v_+$. Since now $[H,S_z]=0$, this ``deformed'' solution
breaks the rotational invariance around the $z$ axis and possesses,
accordingly, a {\it continuous degeneracy}. As seen in Fig.\ \ref{f2}, for
$v_z=0$ $T_c$ is {\it much lower} than $T_e$, lying actually quite close to the
entropic limit temperature $T^s_{e}$. For $b\rightarrow 0$, $T_c\rightarrow
v_+/2$, with $C[\rho(T_c)]\rightarrow 0.55$. Note, however, that for
$b\rightarrow b_0$, $T_c>T_e^d$  due to the logarithmic vanishing of $T_c$ in
this limit, where $C[\rho(T_c)]\rightarrow 1/2$.

{\it 2)} $v_->0$, $v_+=0$. This case of maximum anisotropy ($v_x=-v_y$)
represents, for $v_z=0$, the two-qubit version of the standard Lipkin model,
widely employed in nuclear physics to test symmetry-breaking mean field based
descriptions \cite{RS.80}. It describes the interplay between a single particle
term $bS_z$ and a monopole interaction that induces a deformed mean field. The
states $|\Phi_{1,2}\rangle$ are now entangled, whereas the states
$|\Phi_{0,3}\rangle$ become {\it degenerate}. Hence, in the thermal case
$p_3=p_0$,  and entanglement can only arise from the states
$|\Phi_{1,2}\rangle$, i.e., through the violation of Eq.\ (\ref{i12}) (Eq.\
(\ref{t1}) in the thermal case). This requires $\Delta>-v_z$, i.e., that
$|\Phi_2\rangle$ be the ground state.

The limit temperature $T_e$ determined by (\ref{t1}) depends now on the field
$b$, with $\rho$ entangled for $0\leq T<T_e$ and $C(\rho)$ a decreasing
function of $T$, as seen in Fig.\ \ref{f3} for $v_z=0$. In this case,
entanglement occurs $\forall$ $b$ and
\begin{equation}
 T_e=\Delta/{\rm arcsinh}[{\textstyle\frac{\Delta}{v_-}}]\label{ej6}\,,
 \end{equation}
with $T_e\approx \alpha v_-[1+\half(1-\alpha/\sqrt{2})b^2/v_-^2]$ for
$b\rightarrow 0$. A remarkable feature is that as $b$ increases, $T_e$ now {\it
increases}, even though the entanglement of $|\Phi_2\rangle$ decreases, since
the energy gap $\Delta$ between the ground and the first excited  states
increases. Moreover, for $b\rightarrow\infty$, $T_e\approx
b/\ln(2b/v_-)\rightarrow\infty$, being then possible to make $\rho$ entangled
at {\it any} temperature by increasing the field.
\begin{figure}[t]
\vspace*{-2.6cm}

 \centerline{\scalebox{0.5}{\includegraphics{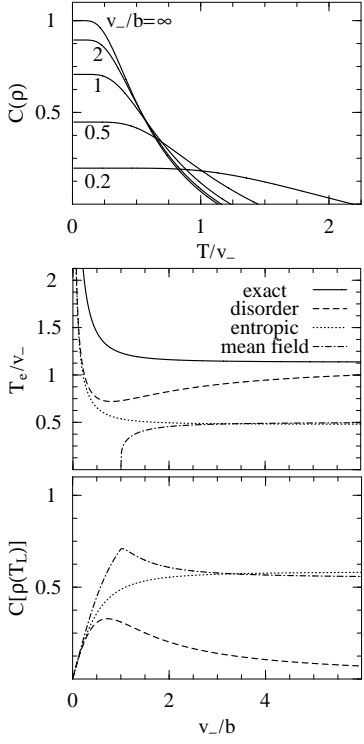}}}
 \vspace*{-2.8cm}

\caption{Same quantities as in Fig.\ \ref{f2} for $v_+=v_z=0$ and different
values of $v_-/b$.}
 \label{f3}
\end{figure}

For $p_0=p_3$, entanglement will be detected by the disorder criterion through
the violation of Eq.\ (\ref{dis}) for $j=1,2$, i.e., of Eq.\ (\ref{dis1}). For
$v_z=0$, this will occur for any value of $b$ but below the lower limit
temperature
 \begin{equation}
 T^d_e=\Delta/{\rm arcsinh}[{\textstyle\frac{\Delta}{\Delta-b}}]\,.\label{ej7}
 \end{equation}
For $b\rightarrow 0$, $T^d_e\approx T_e[1-\alpha b/(\sqrt{2}v_-)]$, with
$C[\rho(T_e^d)]\approx(\sqrt{2}-1)b/v_-$. Eq.\ (\ref{ej7}) is not a monotonous
increasing function of $b$, being minimum at $b\approx 1.25v_-$, but for
$b\rightarrow\infty$, $T_e^d\approx b/\ln [4b^2/v_-^2]\approx T_e/2$, becoming
then also infinite in this limit. Hence, the emergence of entanglement for
large fields is also detected (since it is a ground state effect) but above a
higher threshold. Note also that $T^d_e/T_e\geq 1/2$ $\forall$ $b$, with
$C[\rho(T_e^d)]\leq 0.33$. The limit temperature of the entropic criterion lies
very close to $T_e^d$ for $b\rightarrow\infty$ (as $T^d_e/\Delta\rightarrow 0$
in this limit) but becomes smaller as $b$ decreases, with $T_e^s\rightarrow
0.478 v_-$ for $b\rightarrow 0$.

For $v_z=0$, a phase flip symmetry breaking mean field solution becomes here
feasible {\it only for} $b<b_c=v_-$. For $b>b_c$, ground state correlations,
though non-vanishing, are not strong enough to induce a symmetry-breaking mean
field, so that the entanglement effect for large fields cannot be captured by
the mean field. The permutationally invariant solution corresponds to
$\lambda_x\neq 0$ and $\lambda_y=0$, so that the critical temperature is given
again by Eq.\ (\ref{Temf}) with $v_M=v_-$. Hence, $T_c\rightarrow v_-/2$ for
$b\rightarrow 0$, lying again very close to $T^s_{e}$ in this limit, while
$T_c\rightarrow 0$ for $b\rightarrow b_c$, where $C[\rho(T_c)]\rightarrow
1/\sqrt{2}\approx 0.71$.

{\it 3)} $v_+>0$, $v_->0$. This is the case with {\it finite} anisotropy
$\gamma=v_-/v_+>0$, where entanglement vanishing plus reentry may occur as $T$
increases. For $v_+>v_z\geq 0$, the two lowest states are $|\Phi_2\rangle$ and
$|\Phi_3\rangle$, with $E_2<E_3$ for $\Delta>v_+-v_z$, i.e., $b^2>b_0^2={\rm
Max}[0,(v_+-v_z)^2-v_-^2]$. For $b$ above but close to $b_0$, Eq.\ (\ref{t1})
[(\ref{t2})] will be broken for $0\leq T<T_r^-$ ($T_r^+<T<T_e$), with
$T_r^-<T_r^+$. Hence, as $T$ increases from zero, the concurrence will first
decrease, {\it vanishing} for $T\in[T_r^-,T_r^+]$, but will exhibit a reentry
for $T>T_r^+$, vanishing finally for $T\geq T_e$.

This behavior is depicted in Fig.\ref{f4} for $v_z=0$ and $\gamma=0.7$, where
$b_0\approx 0.71 v_+$ and the reentry occurs for $b_0<b<b_r\approx 1.1 v_+$.
For $b$ close to $b_0$, $T_r^-$ and $T_r^+$ are practically coincident and
equal to the value given by Eq.\ (\ref{Tr}),
$T_r=(\Delta-v_+)/\ln[\Delta/v_-]$, becoming the difference exponentially small
for $b\rightarrow b_0$ ($T_r^+-T_r^-\approx T_r e^{-2v_+/T_r}$). For $b>b_r$
the reentry disappears and $T_e$ becomes the continuation of $T_r^-$,
undergoing then a sharp drop at $b=b_r$. For $b\rightarrow\infty$,
$T_e\rightarrow\infty$, as in case 2, while for $b\rightarrow 0$,
$T_e\rightarrow 0.93 v_+$. At fixed $T<0.93 v_+$, entanglement vanishing plus
reentry will then also occur as $b$ increases.

 \begin{figure}[t]
\vspace*{-2.6cm}

 \centerline{\scalebox{0.5}{\includegraphics{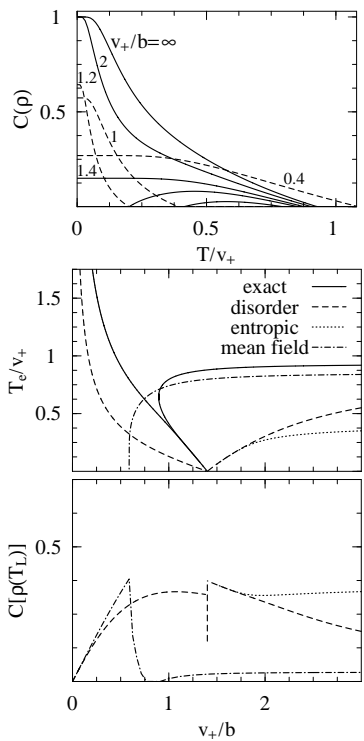}}}
 \vspace*{-2.8cm}

\caption{Same quantities as in Fig.\ \ref{f2} for $v_+>0$ and finite anisotropy
$\gamma=v_+/v_-=0.7$. Entanglement vanishing plus reentry occurs as $T$
increases for $0.9\alt v_+/b\alt 1.4$, as indicated by the solid lines of the
central panel. In the top panel, dashed (solid) lines depict the concurrence in
the interval where Eq.\ \ref{i12} (\ref{i03}) is broken.}
 \label{f4}
\end{figure}
As discussed in sec.\ II, the disorder criterion cannot detect the reentry for
increasing $T$. Instead, the limit temperature $T_e^d$ {\it vanishes} for
$b\rightarrow b_0$, as seen in Fig.\ \ref{f4}, with (\ref{dis1}) broken for
$b>b_0$ and (\ref{dis2}) for $b<b_0$. Nevertheless, $T_e^d\rightarrow T_e/2$
for $b\rightarrow\infty$, so that the entanglement effect for large fields will
be detected, whereas $T_e^d\rightarrow T_e$ for $b\rightarrow 0$, with
$T_e-T_e^d\propto b$ for $b\rightarrow 0$. Note also that $T_e^s$ lies very
close to $T_e^d$ for $b\agt 0.55 v_+$, but becomes lower as $b$ decreases, with
$T_e^s\rightarrow 0.39 v_+$ for $b\rightarrow 0$. Now $C[\rho(T_e^d)]<0.37$
$\forall b$, with $C[\rho(T_e^d)]=0.15$ for $b=b_0$ and
$C[\rho(T_e^d)]\rightarrow 0.37$ and $0.32$ for $b\rightarrow b_0^{\mp}$. This
discontinuity arises from that of $C(\rho)$ for $T=0$ (where $C(\rho)=0.15$ at
$b=b_c$ while $C(\rho)\rightarrow 1$ and $0.7$ for $b\rightarrow b_0^{\mp}$
respectively).

For $v_z=0$, a stable mean field solution breaking phase flip symmetry becomes
feasible only if $b<b_c=v_x$, with $v_x=v_++v_->b_0$ and $T_c$ given by Eq.\
(\ref{Temf}) with $v_M=v_x$. The ratio $T_c/v_+$ is then larger than in case 1.
For $\gamma=0.7$, $T_c$ lies close to $T_e$ for $b\alt v_+$, with
$T_c\rightarrow 0.85 v_+$ and $C[\rho(T_c)]\rightarrow 0.034$ for $b\rightarrow
0$. However, the most striking effect is that $T_c>T_e$ for $1.1\alt b/v_+\alt
1.33$, i.e., for $b$ just above the reentry interval. In this region, $\rho$
becomes separable at a low temperature, yet correlations remain strong to
induce a symmetry-breaking mean field. On the other hand, for $b>b_c$ the
ground state remains entangled but correlations are not strong enough to induce
symmetry-breaking, as occurs in case 2.

\section{Conclusions}
We have examined the exact and the disorder separability conditions for general
mixed states of two qubits interacting through a general $XYZ$ Heisenberg
Hamiltonian, which can be succinctly expressed in terms of total spin
expectation values. The disorder conditions are exact in the absence of a
magnetic field, but become weaker as the field increases and are unable to
detect the reentry of entanglement for increasing temperatures in thermal
states, an effect which may here arise when the ground state is less entangled
than the first excited state. The von Neumann entropic criterion leads to still
lower limit temperatures and is not exact even for zero field. Nonetheless,
both the disorder and entropic criteria do predict the increase in the limit
temperature for large fields occurring in anisotropic models.

The critical temperature for the symmetry-breaking mean field solution is
normally also lower than the exact limit temperature for entanglement in the
examples considered and always vanishes for sufficiently large fields. However,
it can also be higher, particularly when the lowest energy levels are close and
entangled, implying that such solutions, normally regarded as signatures of the
presence of strong correlations in the system, are not rigorous indicators of
entanglement for $T>0$. It is well known that in small systems, the sharp
thermal mean field transitions are to be interpreted just as rough indicators
of a smooth crossover between two regimes. The concept of entanglement allows,
however, to formulate a crossover precisely. Finite systems regain in this
sense a critical-like behavior for increasing $T$, becoming classically
correlated (but not uncorrelated) for $T\geq T_e$, and with an entanglement
undetectable through the eigenvalues of $\rho$ and one of its reductions for
$T_e^d\leq T<T_e$.

{\it Acknowledgments}. NC and RR acknowledge support, respectively, from
CONICET and CIC of Argentina. RR also acknowledges a grant from Fundaci\'on
Antorchas.

\appendix
\section{}
The concurrence of a mixed state $\rho$ of two qubits is a measure of the
entanglement of $\rho$, given by \cite{W.98}
 \begin{equation}
 C(\rho)={\rm Max}[2\lambda_M-{\rm Tr}\,R,0]\,,\label{A1}\\
 \end{equation}
where $\lambda_M$ is the largest eigenvalue of
$R=[\rho^{1/2}\tilde{\rho}\rho^{1/2}]^{1/2}$ and $\tilde{\rho}$ the
spin-flipped density operator, given in the standard basis by
$\tilde{\rho}=(\sigma_y\otimes\sigma_y)\rho^*(\sigma_y\otimes\sigma_y)$, with
$\sigma_y$ the Pauli matrix. The entanglement of formation \cite{Be.96} is an
increasing function of $C(\rho)$ and can be obtained as
 \[{\cal E}(\rho)=-\sum_{\nu=\pm}q_\nu{\rm log}_2q_\nu\,,\;\;
 q_{\pm}=\half(1\pm\sqrt{1-C^2(\rho)})\,.\]
Maximum entanglement corresponds to $C(\rho)=1$, separability to $C(\rho)=0$.
For a pure state $\rho=|\Phi\rangle\langle\Phi|$, $C(\rho)=|\langle
\Phi|\tilde{\Phi}\rangle|$ and ${\cal E}(\rho)$ becomes the von Neumann entropy
of the subsystems \cite{W.98}, $S_2(\rho_A)=S_2(\rho_B)$.

For the state (\ref{4}), the eigenvalues of $R$ are
 \[\lambda_{1,2}=\half\{[(p_1+p_2)^2-
 {\textstyle\frac{b^2}{\Delta^2}}(p_2-p_1)^2]^{\frac{1}{2}}\pm
 {\textstyle\frac{v_-}{\Delta}}(p_1-p_2)\}\,,\]
and $\lambda_{0,3}=p_{0,3}$. Hence, If $\lambda_M=\lambda_{1}$ or $\lambda_2$
($\lambda_0$ or $\lambda_3$), Eq.\ (\ref{A1})  becomes the difference between
the left and right hand sides of Eq.\ (\ref{i12}) [(\ref{i03})] when positive.

The eigenvalues of the partial transpose of (\ref{4}) are
$q_{1,2}=\half[p_{0}+p_{3}\pm{\textstyle\frac{v_-}{\Delta}}(p_2-p_1)]$ and
$q_{0,3}=\half\{p_{1}+p_{2}\pm[(p_3-p_0)^2+{\textstyle\frac{b^2}{\Delta^2}}
(p_2-p_1)^2]^{\frac{1}{2}}\}$, so that the conditions $q_j\geq 0$ $\forall j$
also lead to Eqs.\ (\ref{i}). Only one of them, $q_m$, is negative when $\rho$
is entangled \cite{STV.98}, with $q_m={\rm Min}[q_1,q_2]$ (${\rm
Min}[q_0,q_3]$) if $\lambda_M=\lambda_1$ or $\lambda_2$ ($\lambda_0$ or
$\lambda_3$). In the first case $C(\rho)=-2q_m$ but in the second case,
$C(\rho)\neq -2q_m$ unless $b=0$ or $p_1=p_2$.


\end{document}